\documentclass[superscriptaddress,showpacs,aps,twocolumn,prb,floatfix,nobibnotes]{revtex4-2}
\usepackage{amssymb}
\usepackage{amsmath}
\usepackage{mathtools}
\usepackage{graphicx}
\usepackage{bm}
\usepackage{fontenc}
\usepackage{color}
\usepackage{ulem} 
\usepackage{varioref}
\usepackage{xypic}
\usepackage[hidelinks]{hyperref}
\begin{document}
\title{Topological phase diagram of an extended Su–Schrieffer–Heeger model with negative coupling}

\author{P. A. Grizzi}
\affiliation{Centro At\'{o}mico Bariloche, C.N.E.A, Bariloche, Rio Negro and Consejo Nacional de Investigaciones Cient\'{\i}ficas y T\'ecnicas, Argentina}

\author{P. Roura-Bas}
\email{prourabas@gmail.com}
\affiliation{Centro At\'{o}mico Bariloche, C.N.E.A, Bariloche, Rio Negro and Consejo Nacional de Investigaciones Cient\'{\i}ficas y T\'ecnicas, Argentina}

\author{A. A. Aligia}
\affiliation{Centro At\'{o}mico Bariloche, C.N.E.A, Bariloche, Rio Negro and Consejo Nacional de Investigaciones Cient\'{\i}ficas y T\'ecnicas, Argentina}
\affiliation{Instituto de Nanociencia y Nanotecnolog\'{\i}a
CNEA-CONICET, GAIDI,
Centro At\'{o}mico Bariloche, 8400 Bariloche, Argentina}

\begin{abstract}
Recently, an extended Su–Schrieffer–Heeger model with three atoms per unit cell has been experimentally realized in perovskite exciton–polariton micropillars. A distinctive feature of this system is that the second and third bands undergo a topological transition through a band crossing at the wave vector $k=0$, associated with a negative coupling.
In this work, we analytically determine the phase diagram of the model as a function of the on-site energy splitting and the ratio of intercell to intracell hopping amplitudes. We identify seven regions in this lane, with four distinct topological phases, characterized by different Zak phases of the three bands. Our results provide a useful theoretical framework for interpreting existing experiments and serve as a guide for future experimental investigations of this system.
\end{abstract}

\maketitle

\section{Introduction}
\label{intro}

Topological materials have attracted considerable interest in recent years \cite{Ryu10,Qi11,Ando13}. Experimentally, topological states of matter have been realized in a variety of platforms, including photonic lattices \cite{Recht13,Lu14,ElHassan19,Li24}, scanning tunneling microscopy \cite{Mishra21,Zhao24,Edens26}, ultracold atomic systems \cite{Atala13,Lohse16,Naka16,Vie24}, and polaritonic devices \cite{Jean17,Klembt18,Su21,Wu23,Gagel24,Yu25}.

Several of these studies have focused on realizations of extensions of the Su–Schrieffer–Heeger model (SSHM) \cite{Li24,Jean17,Wu23,Gagel24,Yu25}. In its simplest form, the SSHM describes a one-dimensional dimerized lattice of spinless fermions with alternating nearest-neighbor couplings \cite{Su79}. Originally introduced to describe polyacetylene, the model captures the effects of alternating strong and weak bonds between carbon atoms. It supports distinct topological phases, characterized by a $\mathbb{Z}$ topological invariant, the winding number \cite{Ryu10}. Furthermore, due to the inversion symmetry of the model, the Zak phase is quantized \cite{Asb16}. In particular, for an adequate choice of unit cell, the Zak phase takes the values $0$ and $\pi$ in the trivial and topological phases, respectively \cite{Asb16}, corresponding to winding numbers of 0 and 1.

A variety of extended SSHMs have been investigated in recent years \cite{Li24,Jean17,Wu23,Gagel24,Yu25,Dias22,Aligia23,Ghosh23,Malakar23,Ghuneim24,Joshi25,Chang25,Ali25,He26,Grizzi26,Zwil26}. Some of these extensions support integer winding numbers beyond 0 and 1 \cite{Li24,Dias22,Ghosh23,Malakar23,Joshi25,Chang25,Grizzi26}, while others incorporate three atoms per unit cell \cite{Yu25,Ghosh23}. A particularly interesting recent development is the experimental realization of an extended SSHM with three atoms per unit cell using exciton-polariton micropillars \cite{Yu25}. This platform was designed to realize negative coupling, which reverses the usual energetic ordering of the in-phase bonding and out-of-phase antibonding states. Negative coupling plays a crucial role in several novel physical phenomena \cite{Bena17,Serra18,Bane20} and has also been discussed in the context of biological systems \cite{Stefa08,Liang09}, as outlined in Ref. \cite{Yu25}.

In this work, we derive analytically the exact topological phase diagram of the extended SSHM introduced in Ref. \cite{Yu25}, considering both positive and negative coupling. The paper is organized as follows. In Section \ref{model}, we introduce the model and discuss its symmetries. Section \ref{cross} presents the equations determining the boundaries between topological phases. In Section \ref{topo}, we obtain and analyze the topological phase diagram. Finally, Section \ref{sum} summarizes our results and discusses their implications.

\section{Model and its symmetries}

\label{model}

\begin{figure}[htb]
\begin{center}
\includegraphics[width=\columnwidth]{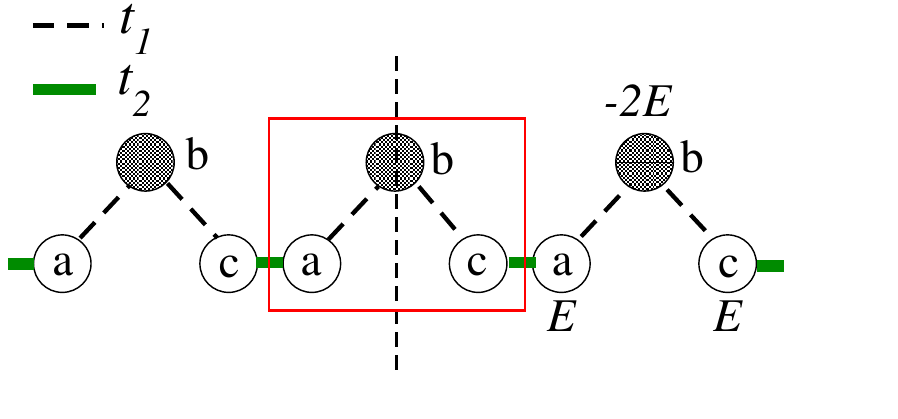}
\end{center}
\caption{Color online. Scheme of the system, with the hopping amplitudes, $t_1$ and $t_2$, and on-site energies, $E$ and $-2E$. The red solid square indicates the choice of the unit cell from which the Bloch Hamiltonian is defined. The vertical dashed line indicates a plane of reflection symmetry.}
\label{scheme}
\end{figure}

A scheme of extended SSHM realized in Ref. \cite{Yu25} is presented in Fig. \ref{scheme}. The Hamiltonian is

\begin{eqnarray}
H &=&\sum_{j}\{E(a_{j}^{\dagger }a_{j}+c_{j}^{\dagger }c_{j}-2b_{j}^{\dagger
}b_{j})  \nonumber \\
&&-t_{1}\left[ (a_{j}^{\dagger }+c_{j}^{\dagger })b_{j}+\text{H.c.}\right]
\nonumber \\
&&-t_{2}\left( a_{j+1}^{\dagger }c_{j}+\text{H.c.}\right) \}.  \label{ham}
\end{eqnarray}
The first term describes the on-site energies of the three atoms per unit
cell, described by the fermion annihilation operators $a_{j}$, $b_{j}$ and
$c_{j}$ for unit cell $j$ respectively. The on-site energies of the first and third sites, $a_{j}$ and $c_{j}$, are equal. For convenience, we set the average of the three on-site energies to zero. The second and third terms
describe the intra- and intercell hopping respectively. Periodic boundary
conditions are assumed.

The Hamiltonian is invariant under time reversal symmetry. Since the model has three bands, particle-hole and chiral symmetries are absent. The solely time-reversal symmetry, which squares to $1$, places the model into the AI (orthogonal) class in the Altland–Zirnbauer (AZ) classification which states that there is no topological non-trivial phases in one dimension \cite{Ryu10}. However, the system is invariant under inversion symmetry with center at any $b_{j}$ site. Therefore, inversion symmetry can still protect a distinct, non-AZ topological phase characterized by a quantized Zak phase \cite{Ryu13}. 

Note that, in addition, the sign of $t_{1}$ can be
changed with the substitution $b_{j}\rightarrow -b_{j}$. Combining this transformation with the simultaneous sign reversal of $E$ and $t_{2}$ maps $H$
onto $-H$.

\section{Eigenvalues and band crossings}
\label{cross}

Taking $t_{1}=1$ as the unit of energy, defining $\alpha =t_{2}/t_{1}$, and
Fourier transforming the operators, the Hamiltonian matrix in the basis of
$a_{k}$, $b_{k}$ and $c_{k}$, for each wave vector $k$ takes the form
\cite{Yu25}

\begin{equation}
H(k)=\left(
\begin{array}{ccc}
E & -1 & -\alpha e^{-ik} \\
-1 & -2E & -1 \\
-\alpha e^{ik} & -1 & E
\end{array}
\right) .
\label{hk}
\end{equation}
The three eigenvalues of this matrix are obtained from
$P(\lambda )=$det$(\lambda \mathbb{I}-H(k))=0$, where $\mathbb{I}$ is the identity matrix. The
determinant becomes

\begin{eqnarray}
P(\lambda ) &=&\lambda ^{3}+\lambda p+q(k),  \nonumber \\
\text{where }p &=&-3E^{2}-\alpha ^{2}-2,  \nonumber \\
q(k) =2E^{3}&+&2E(1-\alpha ^{2})+2\alpha \text{cos}(k).
\label{Pol}
\end{eqnarray}
An analytical solution of the roots of  the equation $P(\lambda )=0$  is
known for centuries. The nature of the roots depend on the sign of

\begin{equation}
D(k)=\frac{q^{2}(k)}{4}+\frac{p^{3}}{27}  \label{dk}
\end{equation}
If $D>0$, there is one real solution and two complex ones. However, since
the matrix $H(k)$ is Hermitian, this case can be disregarded. If $D<0$, the
three roots are real and different. If $D=0$, two real roots coincide
$\lambda _{1}=\lambda _{2}$ and the third one is
$\lambda _{3}=-2\lambda _{1}$. This case is of particular interest because at such points two bands touch, and the topology of these two bands may swap as a function of the parameters. Since $D(k)$ is a smooth function of $k$, and
$D(k)\leq 0$, if $D(k_{0})=0$, then the function $D(k)$ should have a maximum at
$k=k_{0}$. Since $p$ is independent of $k$, it follows that $q(k)$ should have an extremum at $k=k_{0}$.
From $dq(k)/dk=0$ one obtains $\alpha $sin$(k_{0})=0$. Therefore,
for $\alpha \neq 0$, the only possible points at which two bands touch are
$k_{0}=0$ and $k_{0}=\pi $. Note that these two points are protected by the inversion symmetry.

\subsection{Solution for $k=0$}
\label{sk0}

For $k=0$ and $k=\pi $, the states can be classified in even and odd under
inversion symmetry with center at a $b_{j}$ site. The only odd site is $%
|o\rangle =(|a\rangle -|c\rangle )/\sqrt{2}$, where for simplicity the index
$k$ has been dropped. Its energy is $\lambda ^{o}=E+\alpha e^{ik}$. The even
sites are  $|e\rangle =(|a\rangle +|c\rangle )/\sqrt{2}$ and $|b\rangle $.
For $k=0$, projecting $H(k)$ in this basis, the matrix states the form

\begin{equation}
H_{e}(0)=\left(
\begin{array}{cc}
E-\alpha  & -\sqrt{2} \\
-\sqrt{2} & -2E
\end{array}
\right) .
\label{he0}
\end{equation}
The corresponding energies of the even sector are

\begin{equation}
\lambda _{\pm }^{e}=\frac{-E-\alpha }{2}\pm
\sqrt{\left( \frac{3E-\alpha }{2} \right) +2}.  \label{lam}
\end{equation}
The coincidence of the second and third band discussed in Ref. \cite{Yu25}
takes place when $\lambda ^{o}=\lambda _{+}^{e}$. From this equation and the
expressions above one obtains

\begin{equation}
\alpha _{c}(0)=-\frac{3E}{2}+\sqrt{\frac{9E^{2}}{4}+1}  \label{alph}
\end{equation}
for the value of $\alpha$ at which the second and third energies coincide
for $k=0$ and a given value of $E$. In particular, for $E=1/3$, $\alpha _{c}\simeq 0.6180$, in agreement with
the value $\alpha _{c}(0)\simeq 0.62$ reported in Ref. \cite{Yu25} for the
topological transition.

As explained in the next Section, we have verified
that for positive $E$, and positive $\alpha <\alpha _{c}$, the three bands
are topologically trivial, while for $\alpha >\alpha _{c}$, and positive large
enough $E$, the Zak phase of the second and third bands is $\pi $ and
therefore these bands are topologically non-trivial.

Other topological
transitions are described below.

\subsection{Other phase boundaries}
\label{other}

From the form of Eq. (\ref{hk}) it is clear that a shift of the wave vector
in $\pi $ ($k\longrightarrow k+\pi $) is equivalent to a change of sign of
$\alpha $. Therefore from the previous result, one concludes that for
$k=\pi$, the second and third bands coincide for  $\alpha _{c}(\pi )=-\alpha
_{c}(0)$. This case corresponds to positive coupling.

Furthermore, changing the sign of $\alpha $, $E$ and the $b$
states,  $H(k)$ is mapped onto $-H(k)$, as described at the end of Section \ref{model}.
This leads to a crossing between the first and second band at

\begin{equation}
\tilde{\alpha}_{c}(0)=-\frac{3E}{2}-\sqrt{\frac{9E^{2}}{4}+1},
\label{altilde}
\end{equation}
and $\tilde{\alpha}_{c}(\pi )=-\tilde{\alpha}_{c}(0).$

Including the four curves $\alpha _{c}(k)$ and $\tilde{\alpha}_{c}(k)$ for
$k=0,\pi $ leads to the phase diagram shown in Fig. \ref{diag}.

\section{Topological phase diagram}
\label{topo}

\begin{figure}[htb]
\begin{center}
\includegraphics[width=\columnwidth]{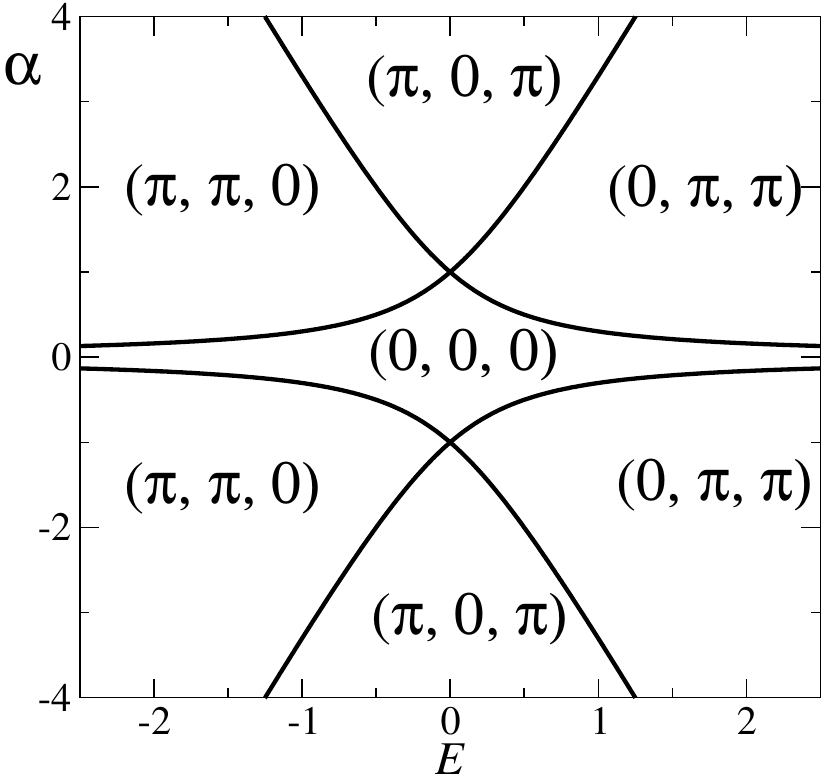}
\end{center}
\caption{Phase diagram of the system characterizing the different topological
phases by the Zak phases of the three bands ordered in increasing energy.}
\label{diag}
\end{figure}

The four curves determined above, corresponding to band-touching points at wave vectors
$k=0$ and $k=\pi$, divide the two-dimensional parameter space
$(E,\alpha)$ into seven distinct regions, as shown in Fig. \ref{diag}. 
The Zak phase of a given band $n$ is given by
\begin{equation}
	\gamma_{n}=i\int_{BZ}\langle u_{n}(k)\vert\partial_{k}u_{n}(k)\rangle,
	\label{zak-phase}
\end{equation}
where the integral runs over the entire $1$D Brillouin zone (BZ) $k\in[-\pi,\pi)$. 
For each region, we have calculated the Zak phases of the three bands for a  discretized BZ  with $N$ points $k_j=-\pi+2\pi j/N$:
\begin{equation}
	\gamma_{n}=-\mbox{Im} \log \prod_{j=0}^{N-1}\langle u_{n}(k_j)\vert u_{n}(k_{j+1})\rangle,
	\label{disc-zak-phase}
\end{equation}
with $\vert u_{n}(k_{N})\rangle=\vert u_{n}(k_{0})\rangle$(periodic boundary in $k$). The result is defined mod(2$\pi)$\cite{Asb16}.
 
The resulting Zak phases, $(\gamma_{1}, \gamma_{2}, \gamma_{3})$, are included inside the corresponding region
in Fig. \ref{diag}. The sum of all Zak phases should be 0 mod(2$\pi)$, because
a completely filled fermionic system should be topologically trivial.
Then, there are four possible distinct topological phases: one in which all three bands are trivial, and three in which a pair of bands have Zak phase $\pi$
while the remaining band has Zak phase 0.
All of these phases are realized in the model for different parameter values.

\section{Summary and discussion}
\label{sum}

We have studied the extended Su–Schrieffer–Heeger model assembled in Ref. \cite{Yu25} using perovskite exciton–polariton micropillars.
We have determined analytically its phase diagram and have identified
seven different regions in the space of parameters which realize the four
different topological phases of the model.

The topological character of the system depends on the filling, which typically takes an integer value. When the three bands are either all empty or all fully occupied, the system is topologically trivial, and a finite chain does not exhibit the edge states expected from the bulk–boundary correspondence \cite{Asb16}.
In contrast, the topological phase investigated in Ref. \cite{Yu25}
characterized by Zak phases $(0,\pi,\pi)$, is nontrivial at filling 2, for which the total occupied Zak phase is $\pi$. Consequently, edge modes are expected to appear in a finite chain at energies within the gap between the second and third bands, in agreement with the experimental observations reported in Ref. \cite{Yu25}. At filling 1, however, this phase is topologically trivial.

The other nontrivial phases identified in Fig. \ref{diag}, exhibit a different dependence on filling: the phase characterized by $(\pi,0,\pi)$ is nontrivial at fillings 1 and 2, whereas the phase characterized by $(\pi,\pi,0)$ is nontrivial only at filling 1.

We believe that our analytical determination of the complete phase diagram provides a useful framework for understanding the system and may stimulate further experimental investigations of related exciton–polariton structures.

\end{document}